# A Trigger And Readout Scheme For Future Cherenkov Telescope Arrays

G.Hermann, C. Bauer, C. Föhr, W.Hofmann, T. Kihm, F.Köck

*Max-Planck-Institut für Kernphysik*
*Saupfercheckweg 1, 69117 Heidelberg, Germany*

**Abstract.** The next generation of ground-based gamma-ray observatories, such as e.g. CTA, will consist of about 50-100 telescopes, and cameras with in total ~100000 to ~200000 channels. The telescopes of the core array will cover and effective area of ~ 1 km$^2$ and will be possibly accompanied by a large "halo" of smaller telescopes spread over about 10 km$^2$. In order to make maximum use of the stereoscopic approach, a very flexible inter-telescope trigger scheme is needed which allows to couple telescopes that located up to ~1 km apart. The development of a cost effective readout scheme for the camera signals exhibits a major technological challenge. Here we present ideas on a new asynchronous inter-telescope trigger scheme, and a very cost-effective, high-bandwidth frontend to backend data transfer system, both based on standard Ethernet components and an Ethernet front-end interface based on mass production standard FPGAs.

## TRIGGERING TELESCOPE SYSTEMS

Whenever the performance of a telescope system is limited either by dead-time of the front-end electronics or by limitations in the data transfer, it is necessary to already suppress on the hardware level background events and to select stereoscopic events. Most of the background events are low-energy hadron showers, muons and accidental coincidences due night sky induced background. As has been shown with the HEGRA, H.E.S.S. and VERITAS telescope systems, combining the telescopes on the trigger level helps to increase the sensitivity of the instruments and to lower their energy threshold.

The main features a central trigger system needs to provide, are as follows:
- Configurable multi-telescope coincidences with arrival time compensation for the shower front depending on the observation direction.
- Short coincidence time of the order of 50 nsec, depending on the field of view and the spacing of the telescopes.
- Support for event synchronization and event building.
- Dead time measurement for the telescopes and the telescope system.
- Monitoring, which telescopes are active or were busy, on event-by-event basis.
- Absolute event timing for phase analysis and multi-wavelength studies.

In addition, a stereoscopic trigger system needs to be flexible and configurable for possible subsystems with different topologies. It needs to be scalable up to large numbers O(100) of telescopes of different types.

So far, the existing inter-telescope trigger systems operate in *synchronous* mode: whenever a camera has generated a (local) trigger, it holds (or buffers) the event on hardware level and sends the trigger information to a central trigger electronics, where the system coincidence condition is checked. In case there is a valid coincidence the telescope gets the information to read and further process the event. Otherwise the event is discarded and/or the camera is reset. Such synchronous inter-telescope trigger systems are highly beneficial if the dead-time due for the readout is much larger than the roundtrip time of the trigger signals from the telescopes to a central trigger station and back. However, in future cameras, due to their high rate capabilities, the front-end dead-time can be as low as a few microseconds or even zero in case of FADC-based solutions. The dead-time in such systems would then be dominated by the round-trip time of the trigger signals to and from the stereoscopic system trigger. In such a case, it is better to first capture events locally, whenever the camera trigger has fired, and to transfer them through a high-bandwidth system into a large buffer memory inside the camera. The selection and pre-processing of stereoscopic events is then done in a second step.

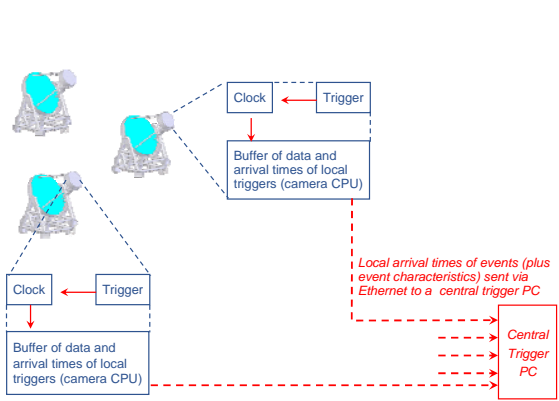

**FIGURE 1.** Ethernet-based scheme for the inter-telescope trigger. The camera computers buffer all events that have been (locally) triggered and send for each event the absolute event time stamps, plus possibly additional event characteristics like the trigger pattern of pixels, to a dedicated central trigger PC. This computer checks for coincidences using the time stamps and tells the camera CPUs which events to keep and which to reject.

In the following we will first show how an asynchronous system trigger scheme could work, and then present an idea for a low cost, high bandwidth data transfer scheme.

## A SOFTWARE BASED STEREOSCOPIC TRIGGER SYSTEM WITH "HARD" TIMING

Figure 1 shows the basic scheme of an asynchronous system trigger. Each camera captures the image data upon a local camera trigger and buffers the data in a multi-event buffer that can keep the data for about a second. At an assumed camera trigger rate of up to 10 kHz and an event size of 60 kByte a buffers size of 1 GByte is sufficient to keep always all events from the last second. With each local trigger an absolute time stamp is captured for the event with an accuracy of the order of nsec and transmitted to the camera CPU. This computer collects the time stamps and possibly additional trigger information for each event, like e.g. pixel trigger patterns, and transmits them every O(10-100 msec) via standard Ethernet using TCP/IP to a dedicated central trigger computer. The central computer receives all time stamps from all telescopes and uses this information to test for time coincidences of the events and to derive the telescope system trigger decision, depending of the user-requested configuration for coincidences and possible sub-systems. In addition the time and trigger information can already be used for a first estimate of the core position and shower direction. Following the central trigger decision, the central trigger CPU sends the information to the corresponding telescopes about which of the buffered events are to be rejected and which fulfill the system trigger condition and should be pre-processed in the camera CPU and transmitted for further stereoscopic processing. Assuming a local trigger rate of each telescope of 10 kHz and about 100 Byte of trigger information coming from each telescope, the central trigger computer needs to handle up to 100 MByte/sec in a 100 telescope system, which can be readily done with today's technology.

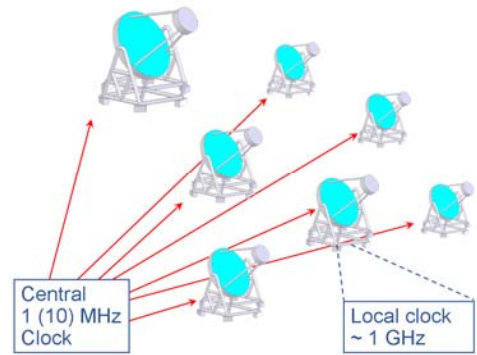

**FIGURE 2.** Clock synchronization scheme for the telescope system.

In such a trigger scheme, the central trigger decision is done in software, while using the "hard" timing from the camera trigger decision. It is therefore scalable, fully flexible and all types of sub-systems can be served in parallel. At the same time it uses the shortest possible coincidence gates and provides an optimum suppression of accidental coincidences. Since the camera CPUs will also send the trigger time information to the central CPU even if the front-end was busy with an earlier event, it will be possible to derive reliable dead-time measurements for the telescopes system as a whole.

One plausible trigger scenario that could be easily implemented in such a software-based scheme requires a coincidence of at least two telescopes in overlapping cells of four telescopes of the system. In addition, if such a condition is fulfilled in at least one cell of the system, the central trigger software could check for coincident events in single telescopes of other cells, and also trigger the further processing of the data of these telescopes. It is expected that with the additional data from such "orphaned" telescopes a further increase in sensitivity in the low-energy domain can be achieved. In case of a hard-wired, "classical" solution, a similar extended trigger and readout condition would require extensive (optical) cabling between cells of telescopes, making such a system expensive and inflexible.

## TIME SYNCHRONIZATION OF THE TELESCOPES

While in traditional, hard-wired central trigger schemes it is not necessary to know the trigger time of individual cameras, in the proposed asynchronous trigger scheme the functionality depends on a proper synchronization of the clocks at different telescopes. In order to obtain the require O(nsec) accuracy, each telescope is equipped with a e.g. 1 GHz relative timing system (like a free running quartz or a TAC system) which is synchronized every O(μsec) through one common central clock, and which is used for the whole system (see figure 2). The signal from the central clock is distributed to all telescopes, either via optical fibers or through a wireless system. Each time the O(μsec) signal arrives at a telescope the GHz timing system is reset and the μsec pulses are counted. Therefore all telescopes share the same time reference and have synchronous clocks with a relative accuracy of O(nsec).

One critical aspect in such a system is the calibration of the relative propagation time of the μsec-pulses to the different telescopes and the monitoring of possible drifts with time or temperature. In case of a fiber-optical system, the initial, crude calibration can be done by measuring the correct length of the fibers with an accuracy of < 1 m, corresponding to ~ 5 nsec. For a more precise determination and monitoring of the relative propagation time, data from cosmic ray events can be used. From the reconstructed core position and shower direction, it is possible to calculate the expected relative arrival time of the light front at different telescopes, depending on the observation direction. By comparing the measured relative arrival time with the expectation, a continuous calibration and monitoring can be obtained. Given the typical expected statistical fluctuations of the relative arrival time of the light front at the different telescopes of a few nsec, an end-to-end precision calibration of the relative timing can be obtained from the data during observations on timescales of less than a minute.

## AN ETHERNET-BASED FRONT-END READOUT SCHEME FOR CAMERAS

The proposed system trigger scheme takes advantage of the fact that in future cameras it will be possible to acquire events with a front-end (or digitization) dead-time of only a few μsec or less. This will be achieved e.g. with the SAM [1] analogue pipeline readout of the H.E.S.S. II telescope, or e.g. with DRS4- [2] or FADC-based solutions. In addition, future digitization systems will provide the option to readout pulse shapes for each pixel, which will be sampled at frequencies of a few hundred MHz up to GHz, depending on the application. In order to make full use of this information, it will be necessary to transmit the digitized pulse shapes of the camera pixels to a camera computer for further processing. Assuming a typical camera with 2000 pixels and 30 Byte of information per pixel and event, a data rate of ~600 MByte/sec needs to be transmitted from the ADCs to a dedicated camera computer with its data buffers. Given the large number of channels in future telescope arrays, a low cost system for high data throughput is needed.

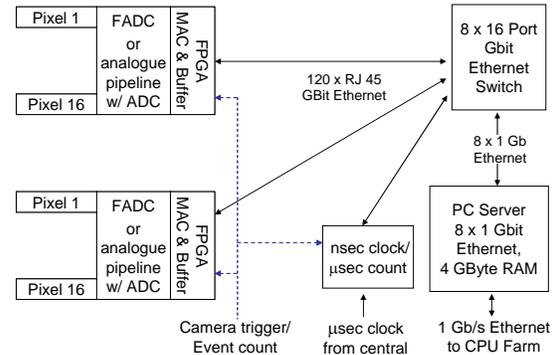

**FIGURE 3.** Possible scheme for an Ethernet-based front-end to back-end readout. A group of pixels with their ADCs is controlled by a dedicated FPGA. The same FPGA can be used to buffer the data and to transmit them through a dedicated Ethernet network to a camera computer ("PC Server"), which buffers the data in its RAM and pre-processes stereoscopic events before sending them to an event building farm.

A possible solution is based on the idea that already the front-end with its ADCs is read out using standard Ethernet components (e.g. on the Ethernet layer 2), which are produced for mass markets and therefore are of low cost. In contrast to "classical" systems of existing cameras, no dedicated synchronous data bus or any type of custom designed data busses are needed. In order to transmit the front-end data, an FPGA based solutions is envisaged. Each group of e.g. 16 or 32 pixels will be served by a dedicated FPGA. This FPGA controls the digitization of the pixel signals either with "slow" ADCs that digitize the signals buffered in analog pipelines, or with fast FADCs. The data from the ADCs is written into a multi-event buffer inside the FPGA and then sent via raw-Ethernet through an Ethernet switch to a camera computer (see figure 3). The camera computer buffers the data until the decision from the central trigger has arrived, and pre-processes and transmits coincident events. The front-end FPGA could be e.g. either a (very) low cost solution like the Xilinx Spartan 3, or e.g. a Xilinx Virtex 5 (or similar devices from other manufacturers), if higher performance is required. Both FPGAs incorporate an Ethernet MAC (or it can be emulated on them) and are therefore suited for this kind of interface. Since the FPGA serves between 16 and 32 pixel, even in case of a hi-end Virtex 5 FPGA the resulting cost per channel will be almost negligible.

Normally the Ethernet network is operated in an asynchronous, non-deterministic way. It therefore might be necessary to tag the events already at the front-end with an event marker. This can be also done on the FPGA, which will use the camera trigger signals to generate event counts that are used for the merging of the data in the camera computer.

One critical aspect in this readout scheme could be the possible loss of Ethernet packets. As long as the loss rate would be less than 0.01 %, in a system with e.g. 100 Ethernet channels the resulting dead-time for the camera would be at a tolerable level of only ~ 1%. However, in a closed system as proposed here, it should be possible to operate in a loss-free mode by profiting from the packet buffering as it is done in today's standard Ethernet switches. First tests using 20 nodes (computers) sending data to one Ethernet interface of a server-PC by using a low-level Ethernet protocol look very promising. Even with low-cost, standard Ethernet switches a loss-free transmission of more than $10^{10}$ packets could be achieved at a total data rate of more than 80 MByte/sec into one Ethernet interface. Current up-to-date servers can operate with 2x4 GBit interfaces and cope with the data. It is therefore expected that a loss-free transmission of the front-end data, even of a 2000 pixel camera operating at data rates of 600 MByte/sec, should not be a problem. Further tests are in progress.

Since the Ethernet system operates in full-duplex mode, it can be also used for the control and parameterization of the front-end components, like HV supplies, trigger settings, digitization settings, etc. It would not be necessary to design a separate command bus, as employed in most current cameras.

## CONCLUSION

We have presented ideas on a new asynchronous trigger scheme for future telescope systems and an Ethernet based readout and control scheme for the front-end of Cherenkov cameras.

Both systems rely on commercial Ethernet components, as they exist already now on the market and are produced in large quantities. It therefore allows for a very low cost and high throughput data acquisition and trigger system, which at the same time can be used to control and monitor the frond-end electronics. It is expected that data rates of up to ~ 1 GByte/sec can be transmitted inside the camera without loss of data. Such a high bandwidth DAQ and trigger system is therefore suitable for very large and medium size telescopes of future telescope systems, as well as for small telescopes, operating at very high energies.

One critical task in the development of this system is certainly the implementation of a transparent Ethernet interface on the FPGA. However, the successful implementation of this interface will provide a universal tool also for possible other applications, like e.g. in high-energy physics or radio-astronomy, but also for industrial applications with high data throughput or in medical instrumentation.